\documentclass{emulateapj}
\usepackage{apjfonts}

\shorttitle{\ion{Fe}{14} recombination rate coefficient}
\shortauthors{E. W. Schmidt et al.}

\begin{document}

\title{Electron-ion recombination measurements motivated by AGN X-ray absorption features:
\ion{Fe}{14} forming \ion{Fe}{13}}

\author{E. W. Schmidt,\altaffilmark{1}
        S. Schippers,\altaffilmark{1}
        A. M\"{u}ller,\altaffilmark{1}
        M. Lestinsky,\altaffilmark{2}
        F. Sprenger,\altaffilmark{2}
        M. Grieser,\altaffilmark{2}
        R. Repnow,\altaffilmark{2}
        A. Wolf,\altaffilmark{2}
        C. Brandau,\altaffilmark{3,1}
        D. Luki\'c,\altaffilmark{4}
        M. Schnell,\altaffilmark{4}
        and
        D. W. Savin\,\altaffilmark{4}
        }


\altaffiltext{1}{Institut f\"{u}r Atom- und Molek\"{u}lphysik,  Justus-Liebig-Universit\"{a}t,
                  Leihgesterner Weg 217, 35392 Gie{\ss}en, Germany,
                  \anchor{http://www.strz.uni-giessen.de/~k3}{http://www.strz.uni-giessen.de/$\sim$k3}}
\altaffiltext{2}{Max-Planck-Institut f\"{u}r Kernphysik,
                  Saupfercheckweg 1, 69117 Heidelberg, Germany,
                  \anchor{http://www.mpi-hd.mpg.de/ion-storage/}{http://www.mpi-hd.mpg.de/ion-storage/}}
\altaffiltext{3}{Gesellschaft f\"{u}r Schwerionenforschung (GSI),
                  Planckstrasse 1, 64291 Darmstadt, Germany,
                  \anchor{http://www.gsi.de/}{http://www.gsi.de/}}
\altaffiltext{4}{Columbia Astrophysics Laboratory, Columbia University,
                  550 West 120th Street, New York, NY 10027, USA,
                  \anchor{http://www.astro.columbia.edu/}{http://www.astro.columbia.edu/}}

\begin{abstract}

Recent spectroscopic models of active galactic nuclei (AGN) have
indicated that the recommended electron-ion recombination rate
coefficients for iron ions with partially filled M-shells are
incorrect in the temperature range where these ions form in
photoionized plasmas. We have investigated this experimentally for
\ion{Fe}{14} forming \ion{Fe}{13}. The recombination rate
coefficient was measured employing the electron-ion merged beams
method at the Heidelberg heavy-ion storage-ring TSR. The measured
energy range of $0-260~\mathrm{eV}$ encompassed all dielectronic
recombination (DR) $1s^2\,2s^2\,2p^6\,3l\,3l'\,3l''\,nl'''$
resonances associated with the $3p_{1/2}\to 3p_{3/2}$, $3s\to 3p$,
$3p\to 3d$ and $3s\to 3d$ core excitations within the M-shell of
the \ion{Fe}{14} ($1s^2\,2s^2\,2p^6\,3s^2\,3p$) parent ion. This
range also includes the $1s^2\,2s^2\,2p^6\,3l\,3l'\,4l''\,nl'''$
resonances associated with $3s\to 4l''$ and $3p\to 4l''$ core
excitations. We find that in the temperature range 2--14 eV, where
\ion{Fe}{14} is expected to form in a photoionized plasma, the
\ion{Fe}{14} recombination rate coefficient is orders of magnitude
larger than previously calculated values.
\end{abstract}


\section{Introduction}

Recent spectroscopic XMM-Newton and Chandra X-ray observations of
active galactic nuclei (AGN) have detected a new absorption feature
around $15$--$17$ {\AA} \citep{Sako2001a, Pounds2001a,Kaspi2002a,
Behar2003a, Steenbrugge2003a,Kaspi2004a,Gallo2004,Pounds2004,
Krongold2005a}. This has been identified as an unresolved
transition array (UTA) due mainly to $2p\to 3d$ inner shell
absorption in moderately charged iron ions with an open M-shell
(\ion{Fe}{1}--\ion{Fe}{16}). On the basis of atomic structure
calculations and photoabsorption modeling, \citet{Behar2001a}
pointed out that the shape of the UTA features can be used for
diagnostics of the AGN absorber. However, \citet{Netzer2003a} noted
a disagreement between the predicted and observed shape of this
feature. As a possible cause for this discrepancy they suggested an
underestimation of the low temperature dielectronic recombination
(DR) rate coefficients for iron M-shell ions. These rate
coefficients determine the charge state balance of the iron M-shell
ions in a photoionized plasma and, consequently, the shape of the
UTA.

Reliable low temperature DR rate coefficients of iron M-shell ions
are not available in the literature. The widely used compilation of
\citet{Arnaud1992} is largely based on theoretical work by
\citet{Jacobs1977} and \citet{Hahn1989a}. The purpose of this early
theoretical work was to produce DR data for modeling coronal
equilibrium. However ions form in coronal equilibrium at
temperatures about an order of magnitude higher than those where
they form in photoionized gas \citep{Kallman2001}. Therefore, it is
questionable to use these theoretical DR data for photoabsorption
modeling. Benchmarking by experiment is highly desirable.

The only DR measurements for M-shell ions of third- and fourth-row
elements available up to now are those for Na-like \ion{Fe}{16}
\citep{Linkemann1995c,Mueller1999c}; Ar-like \ion{Ti}{5} and
\ion{Sc}{4} \citep{Schippers1998a, Schippers2002a}; and Mg-like
\ion{Ni}{18} \citep{Fogle2003a}. Although being of minor
astrophysical relevance, the \ion{Sc}{4} and \ion{Ti}{5}
measurements illustrate that the low-energy recombination spectra of
M-shell ions can be dominated by strong resonances associated with
$3p\to 3d$ core excitations.

This work presents the experimental measurement of radiative
recombination (RR)~+~DR for \ion{Fe}{14} forming \ion{Fe}{13}. In
the temperature range where \ion{Fe}{14} is expected to exist in a
photoionized plasma, our experimentally-derived RR~+~DR rate
coefficient is orders of magnitude larger than the sum of the
recommended RR value from \citet{Woods1981} and the DR value from
the compilation of \citet{Arnaud1992}. As DR is much larger than RR
at these temperatures, this discrepancy is clearly due to errors in
the recommended DR data. A similar discrepancy can be expected for
the other iron M-shell ions. This has important consequences for the
modeling of AGN spectra.

\section{Experimental technique}\label{sec:exp}

The experiment was performed at the heavy-ion test storage ring TSR
of the Max-Planck-Institut f\"{u}r Kernphysik in Heidelberg, Germany
(\mbox{MPI-K}). Measurements employed the well established
procedures for studying electron-ion recombination. Details of
experimental and data reduction procedures as well as further
references are given in \citet{Schippers2001c, Schippers2004c} and
\citet{Savin2003a}.

A beam of $^{56}$\nolinebreak\ion{Fe}{14} was provided by the
\mbox{MPI-K} accelerator facility. After acceleration to an energy
of $4.2~\mathrm{MeV/u}$ the ion beam was injected into the storage
ring where it was collinearly overlapped with a magnetically guided
electron beam. In the overlap region the ions can recombine with the
electrons via RR and DR. Recombined ions were separated from the ion
beam in the first dipole magnet after the electron-ion interaction
region and counted using a single-particle scintillation-counter
with a nearly $100\%$ detection efficiency. The systematic
experimental uncertainty of the measured recombination rate
coefficient is estimated to be $\pm15\%$ \citep{Lampert1996}. This
uncertainty stems mostly from the ion current measurement. Adding
further uncertainties discussed in Sec. \ref{sec:resdisc} increases
this to $18$\%.

The experimental electron energy distribution is best described as a
flattened Maxwellian distribution which is characterized by the
longitudinal and transverse temperatures $T_\parallel$ and $T_\perp$
\citep[see, e.\,g.,][]{Pastuszka1996}. The experimental energy
spread $\Delta\hat{E}=[(\ln(2)\, k_\mathrm{B}T_\perp)^2 +
16\ln(2)\,\hat{E}k_\mathrm{B}T_\parallel]^{1/2}$ depends on the
electron-ion collision energy. Here $k_\mathrm{B}$ is the Boltzmann
constant. In the present experiment the temperatures were
$k_\mathrm{B}T_\perp\approx 12~\mathrm{meV}$ and
$k_\mathrm{B}T_\parallel\approx 0.09~\mathrm{meV}$ and consequently
$\Delta\hat{E}=8.3~\mathrm{meV}$ for
$\hat{E}\lesssim0.2~\mathrm{eV}$, $\Delta\hat{E}=33~\mathrm{meV}$ at
$\hat{E}=1~\mathrm{eV}$ and  $\Delta\hat{E}=245~\mathrm{meV}$ at
$\hat{E}=60~\mathrm{eV}$. In the following the experimental data are
presented as a merged-beams recombination rate coefficients (MBRRC)
$\langle\sigma v\rangle$, i.e., the recombination cross section
times the relative velocity convolved with the experimental electron
energy distribution function.

\section{Results and discussion}\label{sec:resdisc}

\begin{figure}
\figurenum{1} \includegraphics[width=\columnwidth]{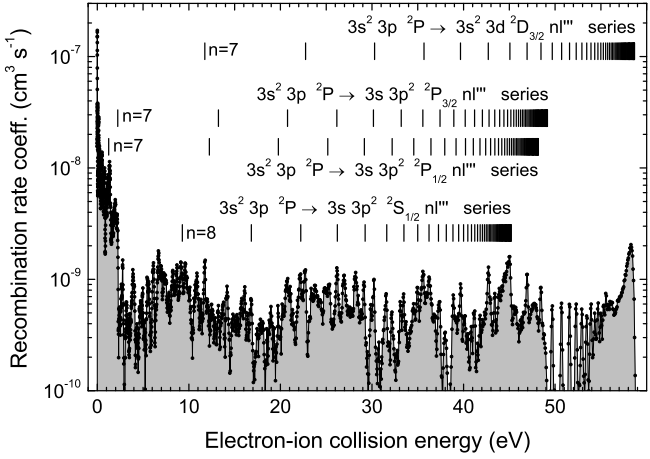}
\caption{Measured \ion{Fe}{14} to \ion{Fe}{13} merged-beams
electron-ion recombination rate coefficient (MBRRC) in the energy
range associated with DR resonances via $3p_{1/2}\to 3p_{3/2}$,
$3s\to 3p$ and $3p\to 3d$ core excitations. The vertical bars denote
DR resonance positions as expected on the basis of the hydrogenic
Rydberg formula. Only the $(3s^2\,3d~^2D_{3/2})\,nl'''$ series of
Rydberg resonances can unambiguously be identified. Note that the
resonances below $2.3~\mathrm{eV}$ exceed all other resonances in
height by an order of magnitude.}\label{fig:Dn0}
\end{figure}

\begin{figure}
\figurenum{2} \includegraphics[width=\columnwidth]{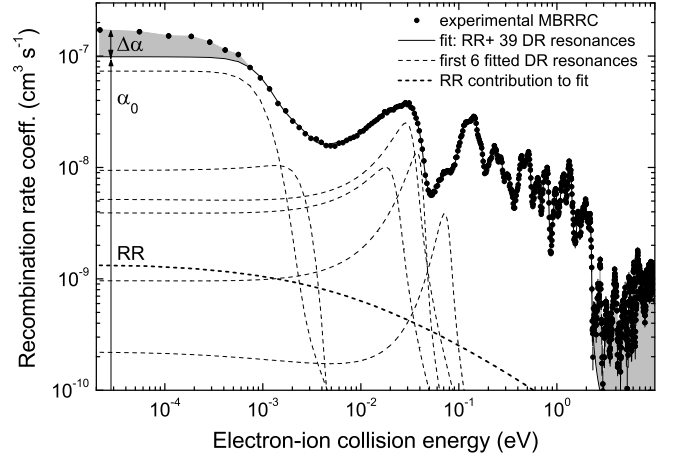}
\caption{Measured \ion{Fe}{14} to \ion{Fe}{13} merged-beams
recombination rate coefficient (MBRRC) at energies below
$10~\mathrm{eV}$ (filled circles). The full curve enclosing the
white area is the result of a fit comprising of 39 DR resonances and
the contribution from RR (short-dashed curve). The dashed curves
show the fits to the first 6 of the 39 DR resonances. The excess
rate coefficient $\Delta \alpha$ (see text) contributes to the
measured signal at energies below $\approx 1~\mathrm{meV}$. Here the
assumed enhancement factor is $(\Delta
\alpha/\alpha_0)+1=1.7$}\label{fig:lowe}
\end{figure}

Figure \ref{fig:Dn0} shows the measured \ion{Fe}{14} MBRRC in the
energy range of $0$--$60~\mathrm{eV}$. This includes all resonances
associated with $3s^2 3p\to 3s^2 3l''$ and $3s^2 3p\to 3s 3p 3l''$
core excitations. The experimental MBRRC increases strongly at
energies below $2.5~\mathrm{eV}$. Figure \ref{fig:lowe} shows this
energy region in more detail. It is dominated by strong DR
resonances that presumably represent levels of the type
$(3s\,3p^2~^2\mathrm{P}_{1/2,~3/2})\,7l'''$. Furthermore the
high-$l$ limit of the $(3s\,3p^2~^2\mathrm{P}_{1/2,~3/2})\,7l'''$
resonances and the series limit of the $3s^2\,3p_{3/2}\,nl'''$
series  at $2.337~\mathrm{eV}$ \citep{Behring1976} are quasi
degenerate within $\approx0.1~\mathrm{eV}$. DR resonances
associated with a $2p^5~^2\mathrm{P}_{3/2}\to {^2\mathrm{P}}_{1/2}$
fine structure excitation were found to be important for
low-temperature DR of \ion{Fe}{18} forming \ion{Fe}{17}
\citep{Savin1997,Savin1999}. The measured \ion{Fe}{18} MBRRC
exhibits a regular Rydberg series of
$(2p^5~^2\mathrm{P}_{1/2})\,nl'''$ resonances. In the present
\ion{Fe}{14} case such a regular resonance pattern from the
$3p_{3/2}\,nl'''$ series is not observed. The question remains to
what extend the unusually strong rise in DR resonance strength
below $2.5~\mathrm{eV}$ is caused by the near degeneracy of
$(3s\,3p^2~^2\mathrm{P}_{1/2,~3/2})\,7l'''$ and
$3s^2\,3p_{3/2}\,nl'''$ resonances. Accurate theoretical
calculations taking into account these features and their mutual
coupling through quantum mechanical mixing appear highly desirable
in order to understand the detailed electronic dynamics leading to
the high resonant rates in this particular energy range.

Up to an energy of $\approx 50~\mathrm{eV}$ the resonance pattern is
irregular. Assignment of the observed features could not be achieved
with the exception of the $(3s\,3p^2~^2\mathrm{S})\,nl'''$ and
$(3s\,3p^2~^2\mathrm{P})\,nl'''$ series limits around $45$ and
$48~\mathrm{eV}$, respectively. At higher energies the
$(3s^2\,3d~^2\mathrm{D}_{3/2})\,nl'''$ Rydberg series is discernable
converging to its limit at $58.6722~\mathrm{eV}$
\citep{Behring1976}. This feature was used to calibrate the
experimental energy scale. This was achieved by multiplying the
experimental energy scale by a factor that deviates from unity by
less than 1\%.

\begin{figure}
\figurenum{3} \includegraphics[width=\columnwidth]{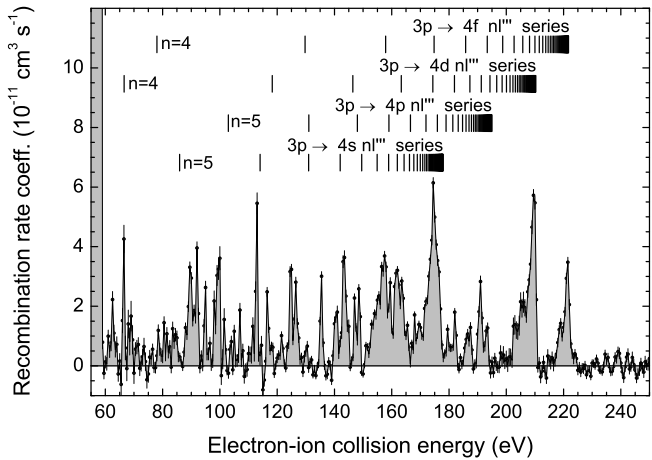}
\caption{Measured \ion{Fe}{14} to \ion{Fe}{13} merged-beams
recombination rate coefficient in the electron-ion collision-energy
region of the $3s^2\, 4l''\, nl''' $ and $3s\, 3p\, 4l''\, nl''' $
resonances attached to the $3p\to 4l''$ and $3s\to 4l''$ core
excitations, respectively.}\label{fig:Dn1}
\end{figure}

Finally, Fig.~\ref{fig:Dn1} shows the measured \ion{Fe}{14} MBRRC
in the range of $60$--$260~\mathrm{eV}$. This range includes
resonances associated with $3s\to 4l''$ and $3p\to 4l''$ (i.e.,
$\Delta N=1$) core excitations. Here $N$ is the principal quantum
number of the transitioning core electron. The strongest features
in this spectral range are the series limits of the $3p\to
4l''\,nl'''$ resonance series. These occur at about
$177.9~\mathrm{eV}~(l''=0)$, $195.0~\mathrm{eV}~(l''=1)$,
$210.4~\mathrm{eV}~(l''=2)$, and $221.7~\mathrm{eV}~(l''=3)$
according the NIST atomic spectra data-base \citep{Ralchenko2005a}.
Identification of the observed peak features with individual
Rydberg series is challenging except for the series limits.

The plasma recombination rate coefficient (PRRC) is derived by
convolving the measured MBRRC with a Maxwell-Boltzmann electron
energy distribution. As detailed by \citet{Schippers2001c,
Schippers2004c}, there are three issues that require special
consideration: the experimental energy spread, the recombination
rate enhancement at low energies, and field ionization of high
Rydberg states in the storage-ring bending magnets.

The experimental energy spread $\Delta \hat E$ influences the
outcome of the convolution for resonances with resonance energies
$\hat E\lesssim\Delta \hat E$. This can be circumvented by
extracting the DR cross section from the measured MBRRC at low
energies. In Fig.~\ref{fig:lowe} this was achieved by fitting $39$
DR resonance line-shapes to the measured spectrum. It should be
noted that most probably none of the $39$ DR line-shapes correspond
to a single resonance but each rather comprises a blend of
resonances. Additionally the fit includes a contribution due to RR
which was calculated using a semi-empirical hydrogenic formula
\citep[Eq.~12 of][]{Schippers2001c}.

An enhanced MBRRC is consistently observed in merged electron-ion
beam experiments at very low energies. For $\hat E\lesssim
k_\mathrm{B}T_\parallel$ the measured MBRRC exceeds the theoretical
expectation by factors of typically $2$--$3$ \citep{Gwinner2000}.
This excess rate coefficient (labeled $\Delta\alpha$ in
Fig.~\ref{fig:lowe}) is an artifact of the merged-beams technique
and hence a mean excess rate coefficient was subtracted prior to the
derivation of the PRRC.

Field ionization of the loosely bound high Rydberg electron in the
recombined ions can result from the motional electric fields they
experience inside the storage ring bending magnets
\citep{Schippers2001c}. In the present experiment only RR and DR
involving capture into Rydberg levels with quantum numbers less than
$55$ contribute to the MBRRC.

Similar to the approach of \citet{Schippers2001c, Schippers2004c}
and \citet{Fogle2005a}, the missing DR resonance strength up to
$n_\mathrm{max}=1000$ was estimated from a theoretical calculation
using the AUTOSTRUCTURE code \citep{Badnell1986}. Although the
AUTOSTRUCTURE code fails to reproduce the irregular \ion{Fe}{14}
resonance structure below $50~\mathrm{eV}$, it reproduces the more
regular resonance structures of high-n Rydberg resonances close to
the various series limits reasonably well when slight `manual
adjustments' are made to the autoionization transition rates.

The unmeasured DR contribution due to $n>55$ was estimated to
contribute ($15 \pm 7$)\% to the experimentally-derived \ion{Fe}{14}
PRRC for plasma temperatures $k_\mathrm{B}
T_\mathrm{e}>9~\mathrm{eV}$. The $7\%$ uncertainty was estimated
from the adjustment of the transition rates. The same field
ionization cut off was taken into account in the RR contribution to
the low energy resonance fit in Fig.~\ref{fig:lowe}. The unmeasured
fraction of the RR-PRRC, due to RR into $3s^2\,3p\,nl$ ($n>55$)
states, was calculated using a hydrogenic formula \citep[Eq.~13
of][Fig. \ref{fig:plasma}]{Schippers2001c}. The RR and DR
contributions from $n=55$--$1000$ were subsequently added to the
PRRC. The contribution of $\Delta N=1$ DR with $n>55$ is
insignificant.

For convenient use in astrophysical modeling codes the total
\ion{Fe}{14} to \ion{Fe}{13} PRRC was fitted using
\begin{equation}
\alpha_\mathrm{plasma}(T_\mathrm{e})=T_\mathrm{e}^{-3/2}\sum_{i=1}^{10}
c_i \exp(-E_i/k_\mathrm{B} T_\mathrm{e}). \label{eq:PRCfit}
\end{equation}
The resulting fitting parameters $c_i$ and $E_i$ are given in Table
\ref{table:PRCfit}. The fit deviates by less than
$1\%$ from the experimentally-derived result in the energy range
$70~\mathrm{meV}-10~\mathrm{keV}$.

Uncertainties in the non-resonant portion of the background that was
not due to RR contributed an estimated 8\% error to the measurement.
Uncertainties in the exact enhancement factor near $0~\mathrm{eV}$
contributed a 2.1\% uncertainty at $0.07~\mathrm{eV}$, and 1.4\% at
$0.1~\mathrm{eV}$, and less than 1\% above $0.14~\mathrm{eV}$.
Adding these, the transition rate uncertainty, and the 15\%
systematic uncertainty in quadrature gives a total experimental
uncertainty of $\pm 18\%$ in the energy range
$70~\mathrm{meV}-10~\mathrm{keV}$.

\begin{table}
\caption{Parameters for the fit of Eq.~\ref{eq:PRCfit}.}
\label{table:PRCfit}
\begin{center}
\begin{tabular}{ccc}
  \tableline\tableline
  $i$ & $c_i$ (cm$^3$ s$^{-1}$ K$^{-1}$) & $E_i$ (eV)\\ \tableline
 1.... & 3.55E-4 & 2.19E-2 \\
 2.... & 2.40E-3 & 1.79E-1 \\
 3.... & 7.83E-3 & 7.53E-1 \\
 4.... & 1.10E-2 & 2.21E0 \\
 5.... & 3.30E-2 & 9.57E0 \\
 6.... & 1.45E-1 & 3.09E1 \\
 7.... & 8.50E-2 & 6.37E1 \\
 8.... & 2.59E-2 & 2.19E2 \\
 9.... & 8.93E-3 & 1.50E3 \\
 10.... & 9.80E-3 & 7.86E3 \\
 \hline
\end{tabular}
\end{center}
\end{table}

\section{Astrophysical implications}

The approximate temperature ranges where \ion{Fe}{14} forms in
photoionized and in collisionally ionized plasmas can be obtained
from the work of \citet{Kallman2001}. For photoionized plasmas they
find that the fractional \ion{Fe}{14} abundance peaks at a
temperature of $4.5~\mathrm{eV}$. From a different photoionization
model, \citet{Kraemer2004a} obtain a value of $7.8~\mathrm{eV}$. The
`photoionized zone' may be defined as the temperature range where
the fractional abundance of a given ion exceeds 10\% of its peak
value. For \ion{Fe}{14} this corresponds to a temperature range of
$2$--$14~\mathrm{eV}$ \citep{Kallman2001}. Using the same criterion
and again the results of \citet{Kallman2001}, for coronal
equilibrium the \ion{Fe}{14} `collisionally ionized zone' is
estimated to extend over a temperature range of
$170$--$580~\mathrm{eV}$. It should be kept in mind that these
temperature ranges are only indicative. In particular, they depend
on the accuracy of the underlying atomic data base.

\begin{figure}
\figurenum{4} \includegraphics[width=\columnwidth]{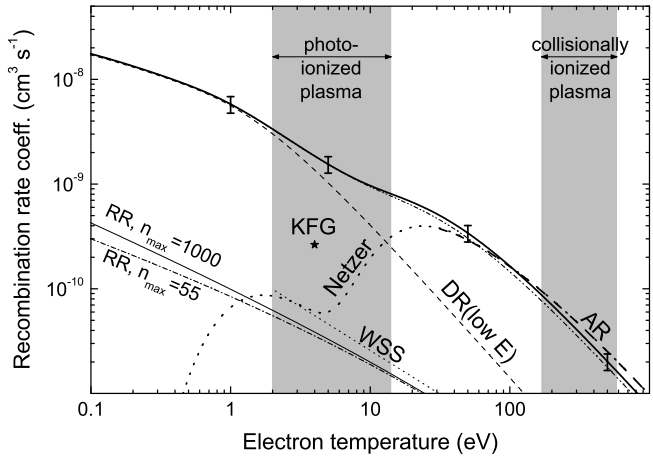}
\caption{\label{fig:plasma} Experimentally-derived \ion{Fe}{14} to
\ion{Fe}{13} recombination rate coefficient in a plasma (PRRC, thick
full line) comprising $\Delta N=0$ DR (Fig.~\ref{fig:Dn0}), $\Delta
N=1$ DR (Fig.~\ref{fig:Dn1}), RR, and the theoretical estimate for
the unmeasured contributions of states with $n>55$ for RR and
$\Delta N=0$ DR. The error bars denote the $\pm 18\%$ experimental
uncertainty in the absolute rate coefficient. The experimental
results without the RR and DR extrapolation are shown by the
dash-dot-dotted line. Also shown is the theoretical DR rate
coefficient of \citet[][thick dashed line, labeled AR]{Arnaud1992},
its deliberate modification by \citet[][dotted line]{Netzer2004a},
and the RR calculation by \citet[][thin dotted line, labeled
WSS]{Woods1981}. The star at $4~\mathrm{eV}$ (labeled KFG) is the
estimate of \citet{Kraemer2004a}. The contribution from the low
energy DR resonances between $0$ and $2.5~\mathrm{eV}$ is shown as
the thin dashed line. The curves labeled RR were calculated using a
hydrogenic formula \citep[Eq.~13 of][]{Schippers2001c} with
$n_\mathrm{max}=1000$ (thin full line) and with $n_\mathrm{max}=55$
(dash-dotted line). The temperature ranges where \ion{Fe}{14} is
expected to peak in abundance in photoionized and collisionally
ionized plasmas are highlighted.}
\end{figure}

In Fig.~\ref{fig:plasma} we compare our experimentally-derived PRRC
with the RR rate coefficient of \citet{Woods1981} and the
recommended DR-PRRC of \citet{Arnaud1992}. This latter is based on a
theoretical calculation by \citet{Jacobs1977} and includes DR
associated with $2p\to 3d$ inner shell transitions. It was
calculated for a temperature range of
$k_\mathrm{B}T=30$--$9\,000~\mathrm{eV}$. In the collisionally
ionized zone the experimentally-derived PRRC is lower than the
theoretical result by $21\%$ at $k_\mathrm{B}T=580~\mathrm{eV}$.
This discrepancy may partly be attributed to the fact that DR
resonances associated with $2p\to 3d$ and higher core excitations
were not measured. But given the poor agreement between the
calculations of \citet{Jacobs1977} and DR measurements for other
iron ions \citep[e.g.,][]{Savin2002c}, this close agreement is
probably fortuitous.

In the photoionized zone the experimentally-derived PRRC is
decisively determined by the resonances occuring at electron-ion
collision energies below $2.5~\mathrm{eV}$. The RR contribution is
insignificant relative to DR. Our derived PRRC is several orders of
magnitudes larger than the DR data from the widely used compilation
of \citet{Arnaud1992}. They are  still about an order of magnitude
larger than the DR-PRRC used by \citet{Netzer2004a} and
\citet{Kraemer2004a} who deliberately assumed sets of DR-PRRC for
the Fe M-shell ions in order to achieve a better agreement of their
plasma modeling results with the observed shape of the Fe $2p\to3d$
UTA (cf.\ introduction).

The present result shows that the previously available theoretical
DR-PRRC for \ion{Fe}{14} forming \ion{Fe}{13} are much too low, as
\citet{Netzer2003a} had already suspected. Other storage ring
measurements show similar deviations from published recommended low
temperature DR-PRRC for M-shell ions
\citep{Linkemann1995c,Mueller1999c,Fogle2003a}. We are now in the
process of carrying out DR measurements for other M-shell iron ions.
As they become available we recommend that these
experimentally-derived PRRC be incorporated into future models of
AGN spectra in order to arrive at more reliable results.

We gratefully acknowledge the excellent support by the MPI-K
accelerator and TSR crews. This work was supported in part by the
German federal research-funding agency DFG under contract no.\
Schi~378/5. DL, MS, and DWS were supported in part by the NASA Space
Astrophysics Research Analysis program, the NASA Astronomy and
Astrophysics Research and Analysis program, and the NASA Solar and
Heliospheric Physics program.

\end{document}